# Measurement of thin films using very long acoustic wavelengths

G.T. Clement, H. Nomura, H. Adachi, and T. Kamakura

The University of Electro-Communications, 1-5-1 Chofugaoka, *Chofu, Tokyo* 182-8585, *Japan. Email: gclement@physics.org*

Short Title: Long wavelength thin film measurement

**Abstract**

A procedure for measuring material thickness by means of necessarily-long acoustic wavelengths is examined. The approach utilizes a temporal phase lag caused by the impulse time of wave momentum transferred through a thin layer that is much denser than its surrounding medium. In air, it is predicted that solid or liquid layers below approximately 1/2000 of the acoustic wavelength will exhibit a phase shift with an arctangent functional dependence on thickness and layer density. The effect is verified for thin films on the scale of 10 microns using audible frequency sound (7 kHz). Soap films as thin as 100 nm are then measured using 40 kHz air ultrasound. The method's potential for imaging applications is demonstrated by combining the approach with near-field holography, resulting in reconstructions with sub-wavelength resolution in both the depth and lateral directions. Potential implications at very high and very low acoustic frequencies are discussed.

**Introduction**

In an effort to resolve objects below the diffraction limit of approximately half the imaging wavelength, a host of so-called superresolution methods have been investigated [1]. Such approaches are generally based on the use of certain *a priori* information about the object to be imaged, thus allowing an optimization routine [2] or a statistical algorithm [3] to surmise higher frequency spatial features. A more objective approach may be applied if the image can be acquired very close to the source [4]. This proximity permits the detection of exponentially decaying and non propagating evanescent waves, which comprise the higher spatial frequencies of the source field. Lateral resolution in these near field techniques is no longer bound to the diffraction limit, but rather is a function of the distance from the source and the dynamic range of the receiver. This range can be realized in optics, for example, through the use of a superlens [5].

In acoustics, such measurements form the basis of acoustic near-field holography [6], which has found applications over a wide spatial scale, ranging from centimeter resolution imaging in industry to 100nm scale microscopy [7]. Although the advantages of such techniques lie exclusively along the lateral plane, perpendicular to the imaging wave, axial measurements are also generally not problematic. Thickness-dependent phase change along the wave direction can be detected at a fraction of a wavelength, whereas attenuation-based amplitude changes can be measured within the limits of the ambient noise level and a given receiver's dynamic resolution. However, the ability to measure either phase or amplitude changes diminishes if the object to be imaged is sufficiently thin. As an object's thickness approaches zero, both its acoustic reflection coefficient and its speed-mediated phase also approach zero, requiring the use of higher frequencies or the ability to achieve high phase resolution for increased sensitivity [8,9]. Such techniques are laudable, but also highly sensitive to environmental fluctuations, limiting their practicality and range of application.

The present study takes the opposite approach. We examine the potential to perform acoustic thickness imaging by *necessarily* using low frequencies – here, using wavelengths up to 90,000X the object thickness. The validity of the method is rooted in momentum transfer through a thin layer. It is well established that the acoustic phase shift experienced by passing through a layer is a function of the layer's density and speed of sound. For most commonly encountered situations, however, density's effect on phase is relatively small as compared to that of the sound speed and propagation distance. As we show here, this is not so under conditions of high density variation over short distances. The present study is an exploration of acoustic wave propagation under such conditions.

**Theory**

The well known acoustic 3-layer problem can be solved by setting the acoustic pressures and normal velocities equal at both sides of the layer boundaries. For a harmonic longitudinal plane wave of frequency f, traveling along the Cartesian z-axis in a medium of sound speed c, density $\rho$, wavenumber $k_0 = 2\pi f / c$, normal impedance $Z_0 = \rho c$, and amplitude $p_0$ incident upon a layer of thickness d and normal impedance $Z_d$, the transmitted longitudinal pressure is given by [10]

$$p(z) = \frac{4Z_0 Z_d p_0 e^{ik_0 z}}{(Z_0 - Z_d)(Z_d - Z_0)e^{ik_d d} + (Z_0 + Z_d)^2 e^{-ik_d d}}, \qquad (1)$$

where $k_d = 2\pi f / c_d$ is the wavenumber within the layer, and $c_d$ is the layer's sound speed. The current work concerns the behavior of the transmitted wave when the layer thickness is much smaller than the imaging wavelength, $\lambda_0 \gg d$, and the layer impedance is much greater than the ambient medium $Z_0 \ll Z_d$. Reduction of (1) under these conditions permits the elimination of small expanded terms above the first order. In this case, the relative phase of (1) upon passing through the layer, reduces to the relation:

$$\phi = \text{Arg}(p(d)/p_o) = \tan^{-1}[d\,(k_0 + k_d \frac{Z_d}{2Z_0})]. \qquad (2)$$

For example, in a thickness range 1 μm to 100 μm, equation (2) would be valid for most common solids and liquids at the audible range of frequencies when air is used as the ambient medium.

Conversely, for thicknesses on an order of a wavelength or greater, the phase of (1) approximately follows its non-reflecting ray approximation,

$$\phi(d) = 2\pi f d \left( \frac{1}{c_d} - \frac{1}{c_0} \right). \qquad (3)$$

A comparison of the phase of (1) and (3) over 10 wavelengths is illustrated in Figure 1a for the exemplary case of a wave in air ($Z_0$ = 410 Rayl) directed through a plastic layer ($Z_d$ = 1.76 X $10^6$ Rayl) whose sound speed ($c_d$ = 1950 m/s) is approximately 5 times faster than air ($c_0$ = 345m/c). Deviation from (3) at longer wavelengths can be seen due to multiple scattering within the layer, but otherwise the lines are in general agreement; the wave is advanced in phase due to the faster speed of the layer. However, this agreement breaks down as d → 0 (Figure 1b) and, in the case of layers much smaller than a wavelength, it becomes apparent that the wave phase actually lags that of a wave traveling a comparable distance through air alone (Figures 1c and 1d),. This is an interesting effect when one considers that the generally faster high-impedance layer would commonly be expected to produce an advanced wave phase. The delay is a consequence of the impulse time as momentum is transferred from a very low density to a very high density.

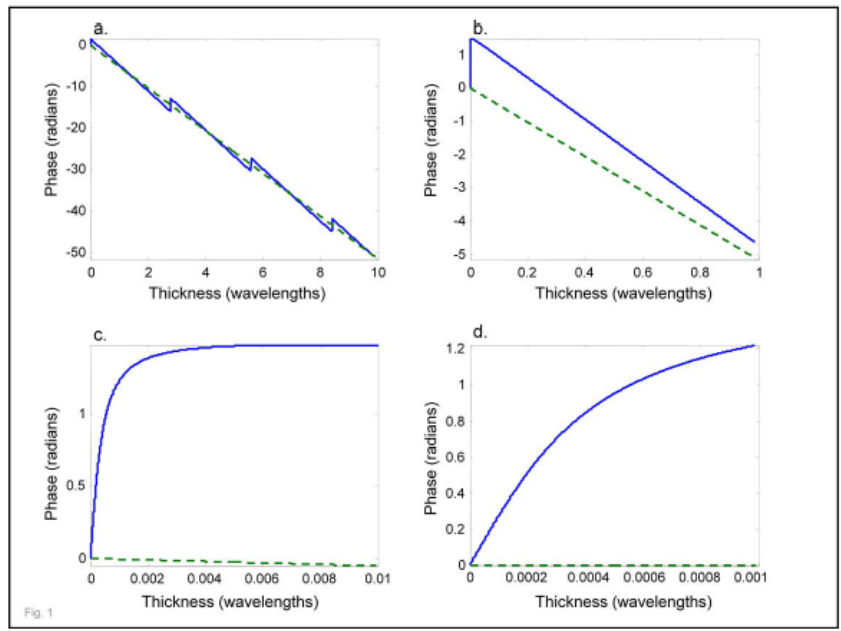

Figure 1. Phase shift as a function of layer thickness through plastic (solid) compared with an idealized calculation using sound speed alone (dashed). Spatial units are given in terms of the imaging wavelength in air. Note functional dependence is dependent on scale as seen over (a) 10, (b) 1 (c) 0.01 and (d) 0.001 wavelengths. The phase shift direction on a fine scale (c) and (d) is opposite that experienced speed-induced phase shifts (a and b).

Considering also the transmission amplitude over the same scaling, it would be expected that the transmission coefficient approaches unity. As can be seen in Figure 2, a significant percent of the wave's energy is transmitted over the same thickness scale where the phase lag becomes appreciable. At such small thicknesses, the transmission coefficients of even very highly attenuating materials approach one.

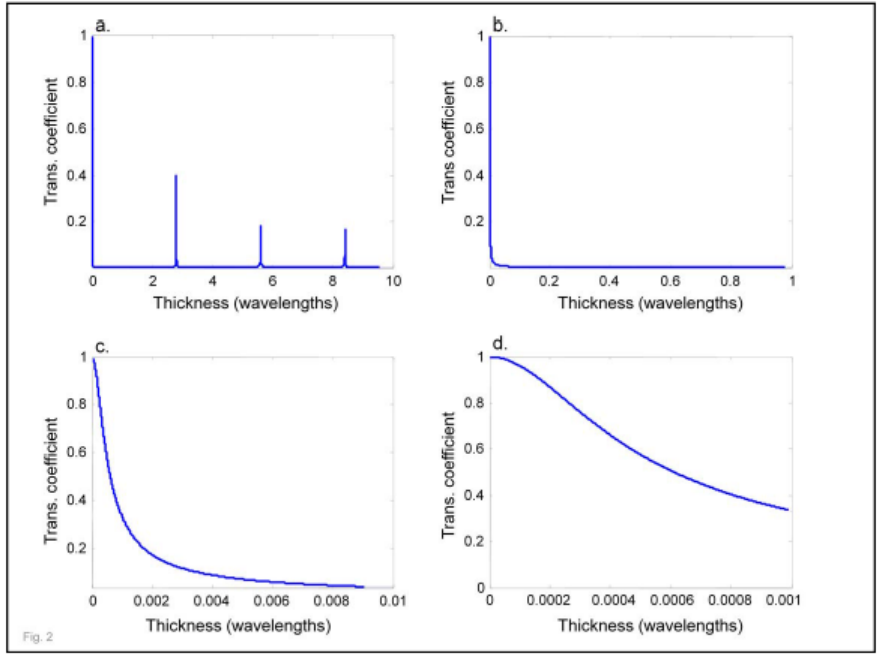

Figure 2. Calculated transmission coefficients corresponding to the phase plots in Figure 1, illustrating shape change as a function of scale. Over multiple wavelengths (a) spectral selection peaks are observed whereas below a wavelength (b) transmission is only appreciable as the thickness approaches zero. This shape transitions toward a 1/r type decay (c) and then toward unity (d).

While the phase curve has an upper bound of $\pi/2$, as can be seen by examination of (2), the actual curvature is primarily determined by the ratio between the layer impedance and the impedance of the ambient medium (Figure 3). Thus, for a given density at a fixed frequency, thickness resolution, R, is set by the phase resolution of the detector. With this observation, the conditions on d are more completely stated as $\lambda_0 >> d \geq R$, with a maximum phase shift of a quarter wavelength at any frequency. In this respect, frequency selection becomes somewhat analogous to the selection of a magnification level in optical microscopy; once the detector resolution is exceeded, a higher imaging frequency can be used.

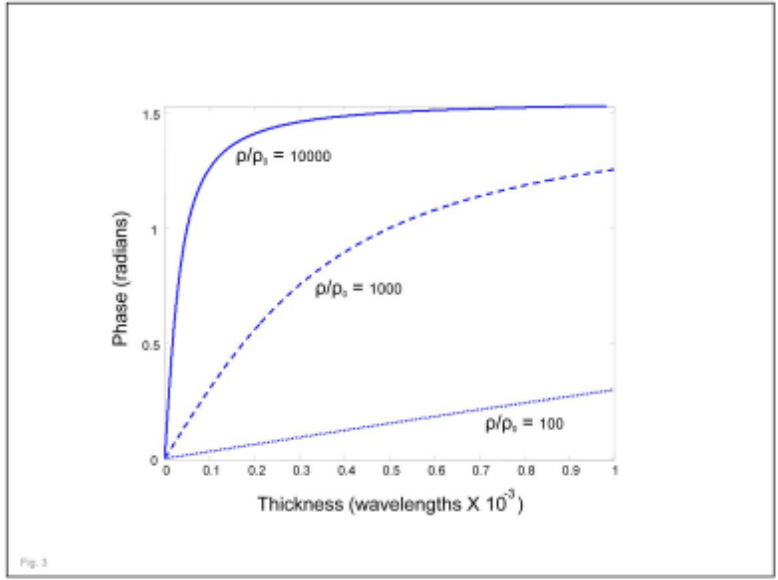

Figure 3. Phase shift as a function of layer thickness in units of wavelengths for 3 different relative densities indicating phase sensitivity increases with density over a scale that is controlled by the wave frequency.

**Methods**

To verify the predicted phase delays, measurements in air were investigated at an audible frequency (7 kHz) through a series of stacked 11 μm polyethylene layers ($\rho$ = 900 kg/m$^3$, c = 1950 m/s) and 25 μm polyester layers ($\rho$ =1180 kg/m$^3$, c = 2540m/s). Samples of 1 to 5 layers were suspended in air using a 29 cm by 20 cm frame and placed between a 20-mm-diameter piezoelectric transducer (Murata 7bb-20-6L0, Kyoto, Japan) and a 10-mm-diameter condenser microphone (Aikoh Electronics CM-102, Tokyo, Japan). Mounts for the transducer, microphone, and frame were selected to allow at least two full wavelengths of reception at the microphone without interference from reflected waves (Figure 4, left). A 7 kHz 5-cycle sinusoidal burst was generated by an analog output of a data acquisition (DAQ) device (X-Series 6366, National Instruments, Austin Texas) and powered by a power amplifier (HAS-4015, NF Corporation, Yokohama, Japan) to supply a 30V output. This output signal was gated to an analog input channel on the DAQ at a 600 kHz sampling rate. Forty waveforms were acquired and averaged for each measurement. Both the polyethylene and polyester measurements were repeated 10 times for analysis. Phase was determined for individual datasets from the phase of the driving frequency. These values were acquired from the Fourier transform of the first three arriving cycles using a timescale referenced to the trigger of the outgoing signal.

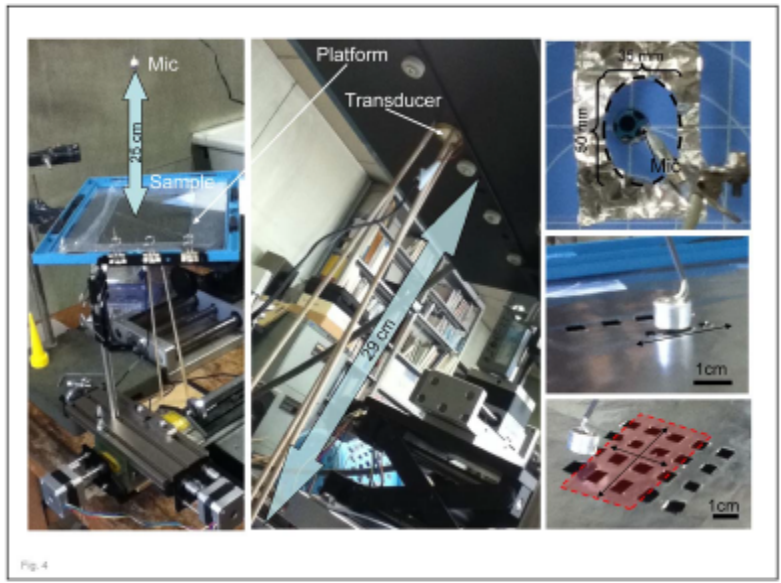

Figure 4. (Left) Images of the setup showing a sample placed above the platform (black) used for imaging. This platform was removed for thickness measurements. (Center) The view below the platform showing the standoff distance of the transducer. The transducer is directed upward at the 0.9 mm aperture. (Upper Right) Image of a soap film measurement. The transducer focal beam arrives through a diaphragm (blue) and is directed toward the microphone. (Middle Right) A 1-dimensional step phantom and (Lower Right) the two-dimensional phantom. The shaded region represents the approximate scan region.

Acoustic pressures through soap films were acquired to verify the method for layers on the micron- and sub-micron scale. To form the films an aluminum frame with an elliptic window (major axis 50 mm, minor axis 35 mm) was inserted into a pan of commercial soap solution (Shabondama, Daiso Industries, Hiroshima, Japan), then oriented vertically, causing film thickness to vary as a function of time (Figure 4, upper right). The film was placed orthogonal to the axis of symmetry of a spherically curved focused air transducer (radius of curvature, 90 mm, diameter 170 mm, assembled in house). The transducer was driven by a 7-cycle 41 kHz sinusoidal wave generated following the same methodology as the 7 kHz measurements. Since the ultrasound field at the transducer's focal plane was approximately 5 mm in diameter (full width at half maximum), waves could be directed through the film without interference from the film's frame. Signals were received by a 3.18 mm-diameter broadband microphone (Brüel & Kjær 4138, Nærum, Denmark), and sent through an amplifier (Brüel & Kjær Nexus) before being recorded by the DAQ at a 2 MHz sampling rate. Waveforms were acquired at 2.5 second intervals. Over the period of acquisition a white light source was directed toward the film to facilitate qualitative observations regarding the color patterns near the ultrasound focus [11] and the time of film rupture. Twenty measurements were acquired for analysis. Phase was determined using the same procedure used as with the 7 kHz measurements described above. Thickness was estimated by correlating the phase against approximate theoretical values ($\rho = 1000$ kg/m$^3$, c = 1500 m/s) using equation (1).

In order to examine the potential to combine thin film density-based measurements with near-field holography, linear and planar measurements were acquired using step phantoms. A 2-step polyethylene step phantom ( 10 mm layer on top of a 20 mm layer on top of a background layer all layers were 11μm thick. See Figure 9, Top), a 1-step aluminum grid (four 2.5 mm openings spaced at 2.5 mm, 12μm thickness) and a 1-step planar aluminum grid (5mm X 5mm openings spaced at 5 mm intervals, 12μm thickness) were used for these measurements. (Figure 4, lower right).

To minimize the effective source size, the 7 kHz emitter was positioned below apertures located on a 10 mm neoprene platform mounted over a 2 mm steel plate (Figure 4, center). Apertures of 3 mm and 0.9 mm were investigated. To prevent vibration coupling between the sample frame and the platform, the frame was mounted such that it floated approximately 1 mm to 2 mm above the platform surface. The microphone was then placed a distance of 1 – 3 mm above the sample. One-dimensional measurements were acquired at 1 mm steps using a 3mm aperture and at 0.5 mm steps using the 0.9 mm aperture.

Linear and planar motion was achieved by affixing the phantoms to a 2-dimensional positioning table (KT 70, Proxxon, Föhren, Germany) retrofitted to function as a stepping motor positioning system using controller (manufactured in house) operated by digital outputs of the DAQ. As signal levels were significantly reduced by the introduction of the aperture, the output voltage was increased to approximately 60V (beyond its specified voltage) and twenty 20 waveforms were acquired and averaged at each location. The two-dimensional image was acquired using the 3mm aperture. It is noted that during this 2D acquisition the piezoelectric transducer source was driven to failure, necessitating replacement with an identical element.

## Results

*Thin film measurements*

Upon passing through both the polyethylene and polyester samples, phase delays were observed for the first three to five arriving wave cycles, after which interference from wave reflections produced variable results. In all cases, the phase delay was observed to increase with increasing layer thickness.

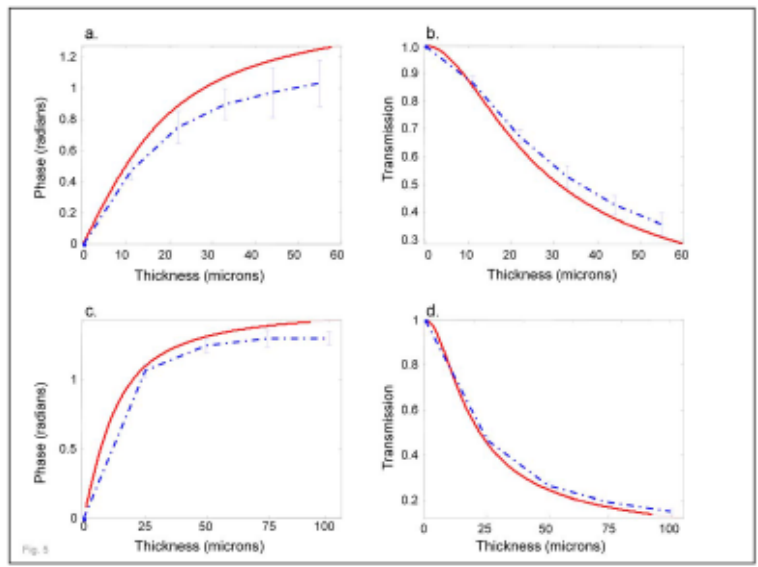

Figure 5. Theoretical and measured data of phase lag due to film thickness for (a) 11 micron polyethylene layers along with (b) the measured and theoretical transmission coefficients. (c) and (d) show similar data for 25 micron polyester layers.

The phases of the acoustic pressure at 7 kHz measured through polyethylene layers are plotted in Figure 5a along with a theoretical curve calculated using equation (1). Phase values are referenced to the pressure recorded with no layers present. The corresponding normalized amplitudes are provided in Figure 5b. Error bars representing two standard deviations are presented with connecting lines passing through the mean values. The data are observed to contain a systematic error that increases with sample thickness, indicating potential error in the layer orientation, or the density and thickness values used in the theoretical calculation. Similar data from measurement through polyester layers are shown in Figure 5c and 5dalong with their corresponding calculated curves.

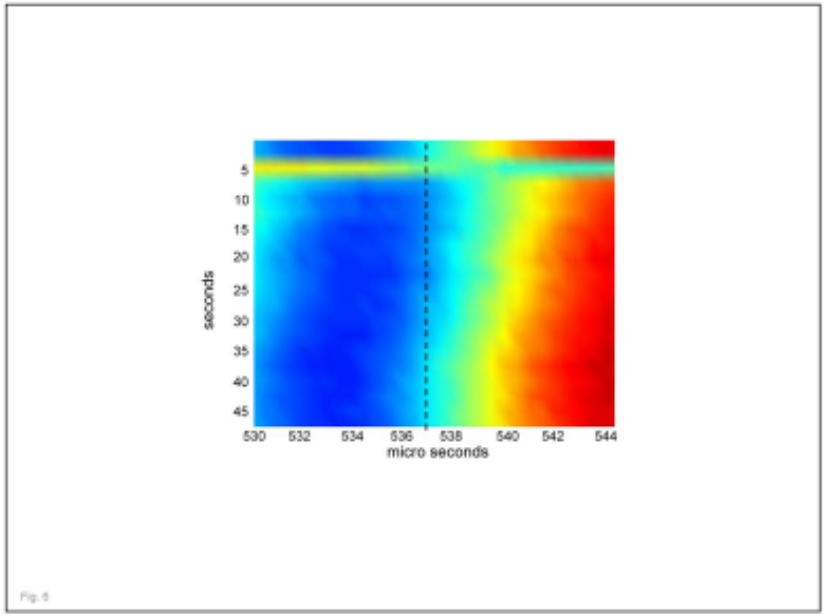

Figure 6. A magnified view of a waves acquired after passing through a soap films. The film is inserted at 5 seconds causing a phase lag that returns to its baseline value at approximately 40 s.

Insertion of a soap film into the focus of a 40 kHz signal was found to cause a phase delay that receded as a function of time, indicative of gravity-induced thinning of the film. The process is illustrated in Figure 6, showing a magnified view of recorded wave cycles. The wave is traveling left to right while the vertical direction represents successive acquisitions. At 5 seconds, the film was inserted causing the initial time lag. As the film thinned due to gravitational draining the wavefront was observed to return to its original phase, indicating the film's thickness had fallen below the detectible resolution set by the frequency and the detector resolution. Data-derived estimates of the initial film thickness of the 20 measurements ranged from 0.8 μm to 1.2 μm, with a mean value of 1.1 μm. In most cases, the bubble ruptured in the time range of 15s to 30s (Figure 7a). In all cases of breakage, a sudden shift in the signal phase to baseline levels was observed. This effect is apparent in the thickness estimates plotted as a function of time in Figure 7b.

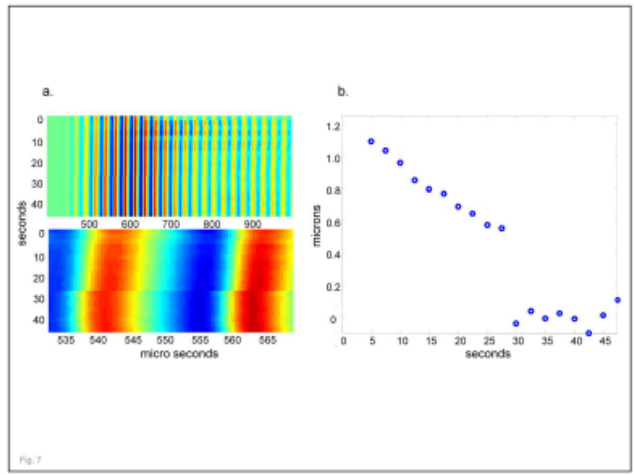

Figure 7. (a.) A view of the wave cycles recorded through a bubble. The close-up view shows the sudden phase shift experienced when the bubble breaks. (b.) The predicted film thickness as a function of time based on phase analysis of the data.

In 4 of the 20 cases very thin films were achieved, as qualitatively evidenced by the transitioning from a color optical interference pattern at the acoustic focal point to an optically black band arriving from the upper portion of the film. This black band is indicative of destructive interference that occurs between outer-surface (phase shifted by pi) reflection and inner surface reflection when thickness-induced phase differences become negligible (< 10 nm) [11]. In such cases, the film was still intact below the resolution set by the selected frequency, with an estimated cutoff resolution of approximately 100 nm (Figure 8).

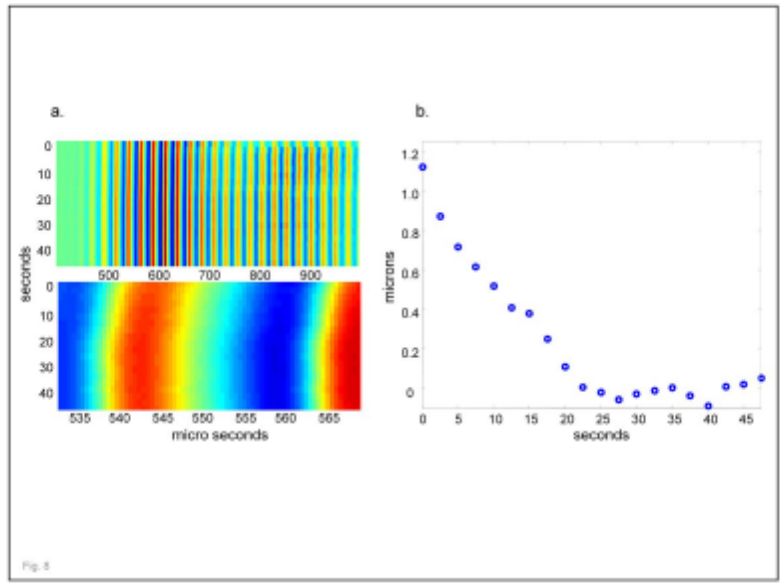

Figure 8. (a.) A view of the wave cycles recorded through a bubble that remains intact over the scale set by the frequency (b.) The predicted film thickness as a function of time based on phase analysis of the data.

*Imaging demonstration*

One dimensional scans over the polyethylene step phantom validated the ability to detect phase shifts along the thickness of the sample at sub wavelength lateral resolution. A magnified view of a time domain wavefront and its associated phase-derived thickness are shown in Figure 9. Phase referenced to the base layer polyethylene. A similar plot showing measurements acquired using the aluminum phantom is provided in Figure 10. In this case the spacing between the steps fail to return to the baseline value of approximately .13 radians, however individual openings are clearly discernable. A two dimensional image formed by scanning over the aluminum grid is shown in Figure 11. The axes indicate the wavelength fraction of the image, indicating features of at least $\lambda/10$ are discernable, while the image intensity indicates thickness variation of $\lambda/4000$ (A horizontal artifact seen in the image is a result of slight movement following breakage and subsequent replacement of the piezoelectric transducer.).

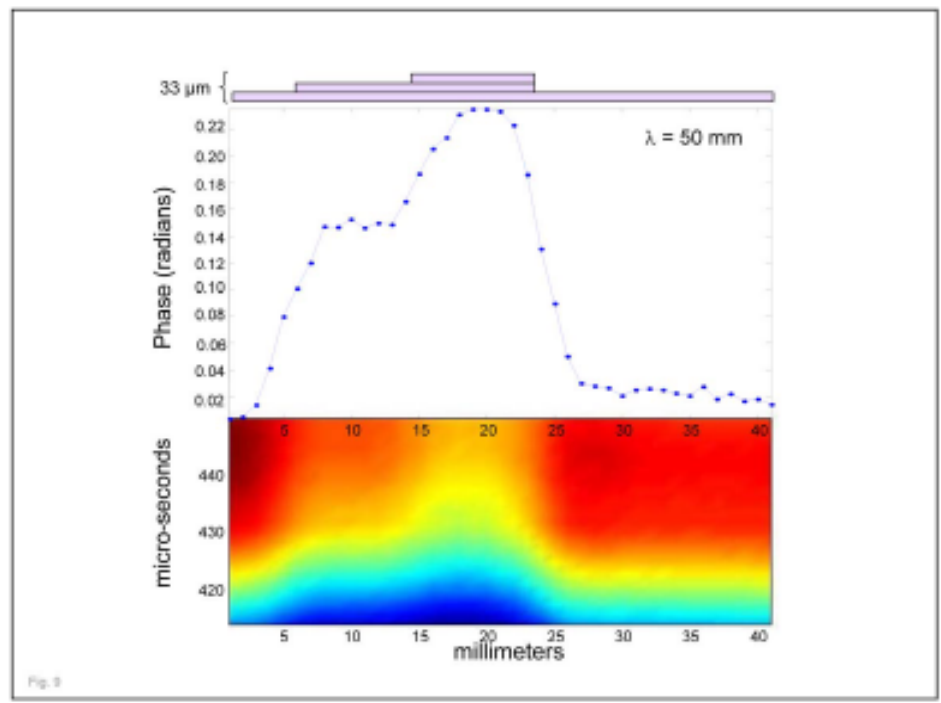

Figure 9. Phase data acquired along a polyethylene step phantom along with an accompanying close-up view of the acquired data.

**Discussion**

In this study, density dependent acoustic phase shifts through thin films were predicted theoretically and verified experimentally. Polyester and Polyethylene films were both observed to induce a phase delay which increased with thickness. Thickness measurements of a soap film, predicted to be as thin as $\lambda/87,000$, were resolved over time. To demonstrate imaging potential the effect was also combined with near field acoustic holography to produce images with a radial resolution of approximately $\lambda/50$ and a thickness resolution of $\lambda/4000$.

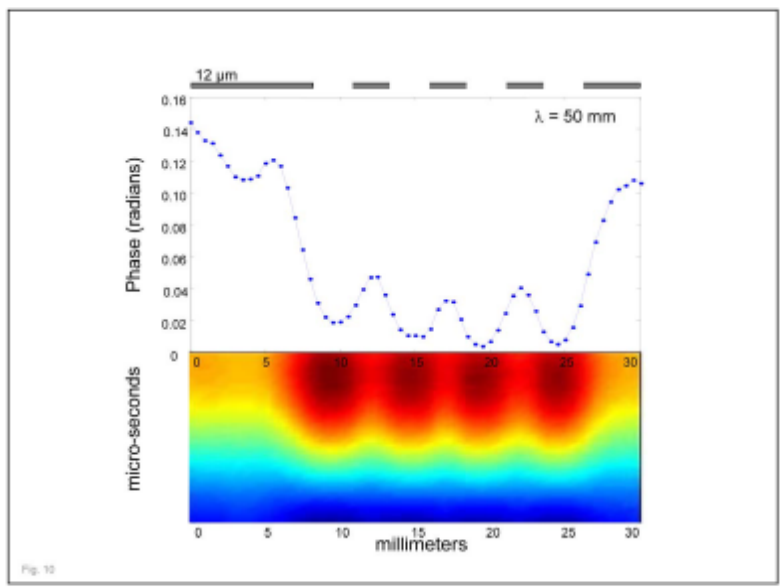

Figure 10. Phase data acquired along an aluminum step phantom along with an accompanying close-up view of the acquired data.

With the exception of the soap film measurements, data were acquired using a very low cost transmitter and receiver, both acquired for the equivalent of a few dollars (USD). The ability to make small-scale measurements with such crude instruments is a testament to the robustness of the approach. The use of long wavelengths made the signals very insensitive to phase fluctuations caused by vibration or thermally-induced variation in the air sound speed. On the other hand, lateral resolution in imaging was compromised by the low output power of the source and the size and dynamic range of the receiver. A higher intensity source and a more sensitive and smaller receiver would be expected to greatly improve lateral resolution.

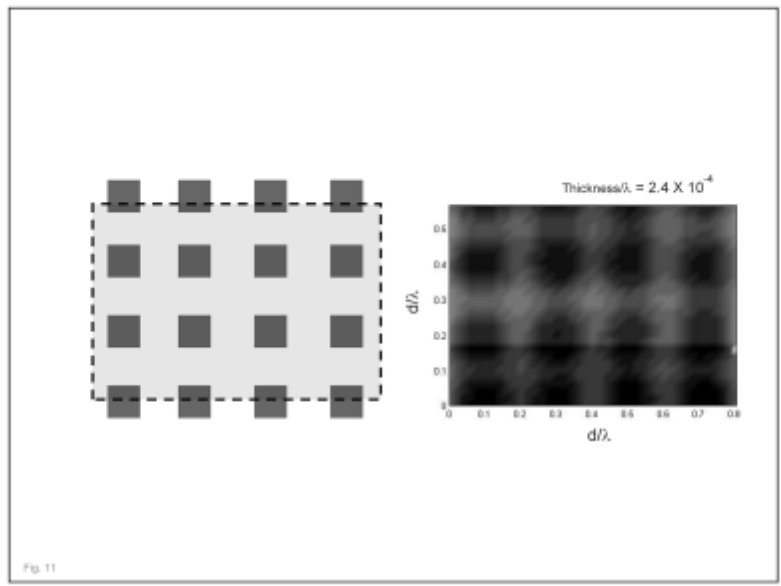

Figure 11. An image of a 12-micron-thick aluminum step phantom acquired at 7 kHz.

By using a very long wavelength, the transmission coefficient through even very highly attenuating and/or highly absorbing materials is high. Since the majority of the acoustic energy is propagated through the material, very little is deposited into the material itself, signifying the approach may be particularly suitable for noninvasive imaging of sensitive biological samples in vitro or in vivo. Furthermore, the approach is readily applicable to metals; a barrier to optical methods.

While the use of audible sound to detect micron-level film thickness makes an motivating demonstration, the importance of the approach likely lies in its implications at very high and very low frequencies. Commercially-available air transducers can generate ultrasound of over 10 MHz in air. At this frequency, resolution below a nanometer would be predicted. Whether this is the case remains to be tested, but clearly the technique would provide an interesting test of mechanical principles applied down to an atomic scale.

Practical and interesting implications may also lie at the low frequency limits of sound. What constitutes a *thin film* is relative to a particular wavelength. On the infrasound scale, objects on the order of a meter might be regarded as thin permitting, for example, non contact measurement through walls, or even the human body. Practical applications would be dependent upon the ability to perform near field holography on a very small scale; clearly a serious engineering challenge, but not a physical limitation.

**Acknowledgements**

This work was supported by The Center for Industrial and Governmental Relations, The University of Electro-Communications.